\documentclass{aa}
\usepackage{graphicx}
\usepackage{epsfig}
\usepackage{lscape}

\usepackage{txfonts}
\begin{document}

\title{The age of extremely red and massive galaxies at very high redshift}
\subtitle{}

\author{N. Castro-Rodr\'\i guez
        \inst{1,2}
	and M. L\'opez-Corredoira
	\inst{1,2}}  
\offprints{ncastro@iac.es}

\institute{
$^1$ Instituto de Astrof\'{i}sica de Canarias (IAC), C/ V\'{i}a L\'actea,
     S/N, Tenerife, E38200, Spain\\
$^2$ Departamento de Astrof\'\i sica, Universidad de La Laguna,
     E-38206 La Laguna, Tenerife, Spain}

\date{Received 2011 June 6; accepted 2011 October 20}

  \abstract
   {}
   {We present a determination of the intrinsic colors and ages of galaxies at very
   high redshift, in particular old galaxies (OGs) within extremely red objects (EROs). To date, 
   the definition of EROs has been restricted to objects with $z<2.5$, however, here consider 
   objects with the same properties but shifted to higher redshifts ($z> 2.5$). We therefore, refer to these objects as very 
   high--redshift EROs (Z-EROS, herein).}
   {We analyze 63,550 galaxies selected in the XMM-LSS field. To obtain a reasonably sized sample of EROs, it 
   is essential to consider a very wide area surveys. We identify targets within an area of 0.77 square degrees for which
   optical to mid-infrared data are available from SUBARU, UKIDSS, and Spitzer. We select Z-EROs
   based on their colors, and then perform a selection of only OGs. One of our novel innovations 
   is to adapt the traditional method of EROs selection based on the filters I and K, 
   to higher redshifts. Using our method, we identify 20
   objects that satisfy the conditions required to be Z-EROs/OGs at redshifts $2.5 \le z \le 3.8$ 
   within some conservative constraints of errors in the photometric redshifts. For each of these objects, we calculate the 
   corresponding color at rest (B-V), and estimate their average 
   stellar mass and age by comparing this color with a synthesis model.}
   {Our selected galaxies have high stellar masses ($\sim 10^{11}$ M$_{\odot}$) and are older than 1 Gyr, hence their stellar populations were formed at z $\gtrsim$ 4.7. After including
   additional galaxies with $z<2.5$ analyzed in a previous paper, we find that the formation epoch depends significantly on
   the observed redshift and stellar mass $\left \langle \frac{\partial t_{\rm form}}{\partial t_{\rm obs}}= -0.9\pm 0.3 \right \rangle$, 
   $\left \langle \frac{\partial t_{\rm form}}{\partial\log _{10}M_*}=-4.8\pm 1.8 \right \rangle $ Gyr. That is, the higher the
   stellar mass, the lower the age of the Universe at which it was formed. This result appears to conflict with
   $\Lambda$--CDM models that claim that the most massive galaxies formed after lower mass.}
   {}


\titlerunning{Ages of red galaxies at high redshift}
\authorrunning{Castro-Rodr\'{i}guez \& L\'opez-Corredoira}
 
   \maketitle
%


\section{Introduction}

An important goal of extragalactic astronomy is to understand how stars are assembled into galaxies and how this assembly 
is related to their evolution. In hierarchical $\Lambda $--CDM models, star formation starts out in low-mass systems, 
which build more massive galaxies through sequential merging (e.g., White \& Rees 1978; Somerville 2005; De Luc\'\i a 
et al. 2006). In this picture, the most massive galaxies should be found at relatively low redshifts. This is 
apparently at odds with observations of massive galaxies at high redshift: a significant population of galaxies with 
stellar mass $\sim$10$^{11}$ M$_{\odot}$ has been found at z$\sim 2{\mbox{--}}$ 3  (e.g., Fontana et al. 2004; Yan et al. 2004; Daddi et al. 2005; Rudnick et al. 2006; van Dokkum et al. 2006; Cassata et al. 2008; Kriek et al. 2008, 2009; Hempel et al. 2011; L\'opez-Corredoira 2010 [hereafter L10]). Stellar population synthesis models combined with broadband photometric data have been used to show that many of these galaxies contain an old stellar population, with ages that indicate a star formation phase occurred within 1-3 Gyr of the Big Bang.
Even at redshifts higher than 3, there is evidence of massive early-type galaxies (Toft et
al. 2005; Chen \& Marzke 2004; Rodighiero et al. 2007; Wiklind et al. 2008). These intriguing results have motivated efforts to find out 
more about these high redshift massive galaxies, particularly the epoch at which they 
formed, which provides an important check of their compatibility with cosmological models.

It is also often accepted that star formation in galaxies was more active at high redshift and therefore that these 
galaxies are in general intrinsically bluer than at low redshift (e.g., Dickinson et al. 2003; Rudnick et al. 2006; 
Labb\'e et al. 2007), as would be expected if their populations were younger and had lower mass/luminosity ratios. 
However, there are also very red galaxies at very high redshift, it is indeed well-established that 
at intermediate-high redshifts a significant population of red passively evolving early-type old galaxies'' (OGs) 
can be found in the field among extremely red objects (Cimatti et al. 2002; Yan et al. 2004; see McCarthy 2004 for 
a review), together with ``dusty-reddened starbursts'' (DSs). All these galaxies are ``extremely red objects'' (EROs). 
In our case, we are interested not in DSs but in finding old galaxies at high redshift. The analysis of EROs/OGs provides a 
convenient means of achieving this aim because we can derive their ages using simple synthesis models. Since EROs are a mixture of OG and DS 
populations, it is necessary to remove DSs from ERO samples before the OG population can be investigated (Miyazaki et 
al. 2003, Fang et al. 2009).

In this paper, we continue the studies initiated by L10 to determinate the
intrinsic color and age variations at different redshifts in a statistical way for very red passively evolving galaxies. 
While L10 analyzed EROs/OGs with $z<2.5$, we analyze EROs/OGs at higher redshift ($z>2.5$). Since 
the EROs are rare and clustered (Daddi et al. 2000; Roche et al. 2002; Miyazaki et al. 2003; Georgakakis et al. 2005), and even rarer at $z>2.5$, wide-field surveys are essential to help us study their 
statistical properties, hence, in this paper we use the large area survey XMM-LSS.
The paper is organized as follows. In \S \ref{.criteria}, we describe the criteria used to find the candidates to very 
red galaxies. In \S \ref{.data}, we describe our sample and photometric data. In \S \ref{.colors}, we 
calculate the colors at rest of the galaxies. In \S \ref{.age}, we derive the ages of these galaxies. 
And in \S \ref{.disc}, we discuss the results.
Throughout this paper, we assume a $\Lambda$--CDM cosmology with $\Omega_m$ = 0.24, $\Omega_{\Lambda}$ = 0.76, and H$_{0}$ = 73 Km/s/Mpc.

\section{Criteria for selecting extremely red objects/old galaxies at $z>2.5$}
\label{.criteria}

It is possible to use colors to identify EROs/OGs within a sample of galaxies. For instance, among the many references,
Fang et al. (2009) and Pozzetti \& Mannucci (2000) presented methods to select EROs/OGs in the UDF field using magnitudes in the three filters $i_{7750}$,
J, and K. The criterion for selecting EROs is (using always AB magnitudes; Fang et al. 2009).

\begin{equation}
(i_{7750}-K)>2.42
\label{Fang_EROS}
,\end{equation}
and likewise, the criterion to select OGs within EROs is
\begin{equation}
(J-K)<0.20(i_{7750}-K)+0.39
\label{Fang_OGs}
,\end{equation}
which is valid only for galaxies with redshifts $0.8<z<2.5$. There are also other methods for selecting OGs within
ERO galaxies, and Fang et al. (2009) demonstrated that all of them are more or less consistent 
with each other. 
There may be some incorrect identification of OGs among the galaxies selected
with this color method (Fang et al. 2009 estimated the contamination to be $\lesssim 25$\%),
and consequently may be present dusty galaxies in the sample selected using this criterion, 
but the statistical comparison with the SED fitting method in Fig. 1 of L10 
shows that there are neither significant differences nor systematic effects with 
redshift. 

Equations (\ref{Fang_EROS}) and (\ref{Fang_OGs}) for a galaxy at $z=1.5$ (the approximate center of the redshift
interval) are equivalent to

\begin{equation}
\left \langle \frac{d\log F_\nu }{d\log \lambda }\right \rangle (i_{7750},K)>2.22
,\end{equation}
\begin{equation}
\left \langle \frac{d\log F_\nu }{d\log \lambda }\right \rangle (J,K)<0.68+0.38
\left \langle \frac{d\log F_\nu }{d\log \lambda }\right \rangle (i_{7750},K)
,\end{equation}
where the $\langle (...) \rangle $(filter 1, filter 2) expressions indicate the average over the wavelength range 
between the center of filter 1 and the center of filter 2. Moreover, if we translate these equations in to the
equivalent one at higher redshifts (taking the closest AB filter redshifted instead of filters $i_{7750}$, J,
K), we get

\begin{equation}
\begin{array}{ll}
J-[3.6]>2.55,\\ K-[3.6]<0.39 +0.19(J-[3.6]),& \mbox{$2.5<z\le 3.5$}\\
\\
H-[4.5]>2.41,\\ K-[4.5]<0.56 +0.29(H-[4.5]),& \mbox{$3.5<z\le 4.5$}\\
\\
H-[5.8]>3.02,\\ $[3.6]$-[5.8]<0.35 +0.14(H-[5.8]),& \mbox{$4.5<z\le 5.5$}\\
\\
K-[5.8]>2.43,\\ $[3.6]$-[5.8]<0.35 +0.18(K-[5.8]),& \mbox{$5.5<z\le 6.5$}\\
\\
K-[8.0]>3.21,\\ $[3.6]$-[8.0]<0.59 +0.23(K-[8.0]),& \mbox{$6.5<z\le 7.5$}
\end{array}
\label{criteria_eq}
,\end{equation}
where [x] stands for the magnitude at filter of x $\mu$m. These are the conditions the EROs/OGs sources must 
satisfy when red-shifted to $z>2.5$. We refer to these sources as Z-EROs/OGS. We apply this
selection to more than 60,000 galaxies, as we see below. For a more secure selection and 
to avoid contamination by starbursts, we also apply another condition on the ratio of
fluxes in K to [24.0] (Fontana et al. 2009) given by
\begin{equation}
F([24])/F(K) < 6.
\label{criteria2_eq}
\end{equation}
This criterion is valid for redshifts of 2.5-4.0. As explained in Fontana et al. (2009), for quiescent and ``red and dead'' galaxies, 
this criterion effectively applies a threshold to the specific star formation rate by comparing with theoretical predictions.

\section{Data}
\label{.data}

Our main purpose is to extend the sample of L10 to include galaxies at $z>2.5$, which were not analyzed by L10.  We require flux data for at least five filters from B to [8.0],
including always the K filter and then three appropriate filters in the range
required for the two color determination according to the system of equations given in Eq. (\ref{criteria_eq}), with a signal-to-noise ratio (S/N)
greater than 3. We also insist that there is a small amount of emission in the filter [24.0] to apply the selection criterion of Eq. (\ref{criteria2_eq}). If we do not have any signal in that band, we assume that the object has no detectable emission at [24.0]. 

After exploring the data of the different available public surveys, we realized that we needed
to combine a mid-infrared survey with both optical and near-infrared ones covering large areas and deep enough to reach the high 
redshift sources. The survey UKIDSS (Hewett et al. 2006; Lawrence et al., 2007) is perhaps the highest quality survey in near infrared with 
these characteristics. It consists of five surveys of complementary combinations of depth and area, employing the 
wavelength range 0.83-2.37 $\mu$m in up to five filters ZY JHK, and extending over both high and low Galactic 
latitude regions of the sky. The UKIDSS uses the Wide Field Camera mounted on the UK Infrared Telescope (UKIRT) 
(Casali et al. 2007). Our study uses data from the DR3 (Data Release 3) of the Ultra Deep Survey (UDS), specifically the 
XMM-LSS field centered on RA=02h18m00s and DEC=-05d00m00s (J2000), which is a deep, wide survey with an area of 0.77 deg$^2$ 
and a 5$\sigma$ depth up to J $\sim$ 24.9, H $\sim$ 24.2, and K $\sim$ 24.6 in the AB system (Lawrence et al. 2007). 
An early data release of UKIDSS was previously used to study EROs (Simpson et al. 2006).

The optical imaging observations of this field were carried out using the prime-focus camera (Suprime-Cam; Miyazaki 
et al. 2002) on the Subaru Telescope in the period from 2002 September to 2005 September as part of the project 
SUBARU/XMM-Newton Deep Survey (SXDSS; Furusawa et al. 2008). The layout of the pointings is arranged in a cross 
shape so that each of the north-south and the east-west directions has an extent of $\sim$1.3 deg (1.2 deg$^2$). 
This corresponds to a field span of a transverse dimension of $\sim$75 Mpc at z$\sim$ 1  and $\sim$145 Mpc at 
z$\sim$ 3  in the comoving scale for (h,$\Omega _{M}$,$\Omega _{\Lambda }$) =( 0.7,0.3,0.7 ). Observations of the 
SXDF were performed using five broadband filters, B, V, R$_{c}$, i', and z' to cover the entire wavelength range observable 
with Suprime-Cam. The limiting magnitude in each band is B=28.2, V=27.8, R$_{c}$=27.5, i$^{\prime }$=27.2, and 
z$^{\prime }$=26.3  (AB, 5 $\sigma$, $\phi =2^{\prime \prime }$ ) in the deepest images of the five pointings. 

We have the additional advantage that one of the regions covered by UKIDSS and SUBARU was covered by the 
Spitzer Wide-area InfraRed Extragalactic survey (SWIRE; Lonsdale et al. 2004). The SWIRE XMM-LSS field survey has an area 
of 9.1 deg$^2$ in all seven Spitzer bands: Infrared Array Camera (IRAC; Fazio et al. 2004) at 3.6, 4.5, 5.6, and 8 
$\mu$m; and Multiband Imaging Photometer (MIPS; Rieke et al. 2004) at 24, 70, and 160 $\mu$m. The sensitivities in all 
the bands are 3.7 $\mu$Jy, 5.4 $\mu$Jy, 48 $\mu$Jy, and 37.8 $\mu$Jy for the IRAC bands and 230 $\mu$Jy, 18 mJy, and 
15.5 mJy for the MIPS bands. The Spitzer Space Telescope has permitted the selection of objects in a spectral region (3-8 $\mu $m) that was not previously accessible from the ground. 
The selection of objects at these wavelengths is particularly invaluable for high redshift 
massive galaxies because a 3-8 $\mu $m selection allows a sampling of the rest-frame near-IR flux of the high-z galaxy 
spectral energy distributions (SEDs) that depends more strongly on to their stellar mass than their star 
formation activity, and also is much less affected by dust extinction effects. 

\begin{figure}
\begin{center}
\vspace{1cm}
\includegraphics[width=8.cm]{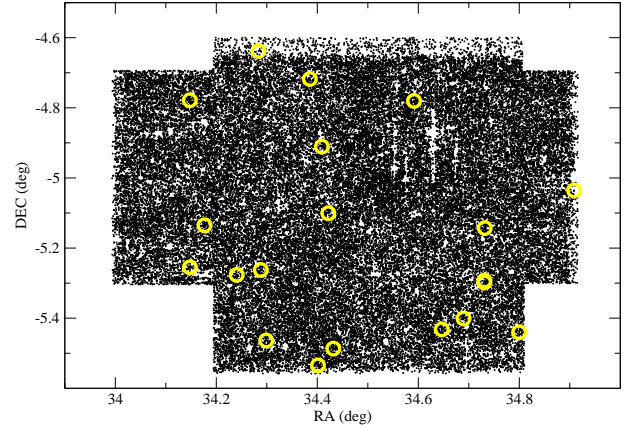}
\end{center}
\caption{Positions of the 63,550 galaxies of the merged catalogs SUBARU, UKIDSS, and Spitzer. The light 
circles correspond to the 20 selected galaxies complying with the criteria for Z-EROs/OGs given in
section \ref{.criteria}.}
\label{Fig:covered_area}
\end{figure}

Therefore, we decided to combine these three surveys to compile a sample from which to extract our sources. 
We determined the objects in common to the UDS DR3+ K-band and the SUBARU catalog from Simpson et al. (2010), and cross-correlated 
this list of objects with the SWIRE data in the XMM-LSS region using the RA and 
DEC positions to within an error of 1 arcsec. The final area covered is 0.77 deg$^2$ (see Fig. \ref{Fig:covered_area}). In the borders of the Fig. \ref{Fig:covered_area}, the deep is lower but its means that we can lose some Z-ERO/OG there (1 or 2). Star-galaxy separation was made with the Bz-K diagram: $ B-i'$ and $ B-K_\mathrm{s}$ colors and the FWHM of objects 
following the criterion by Daddi et al. (2004). In what follows, colors are measured in a 3 arcsec aperture, while 
magnitudes are total magnitudes from SExtractor MAG AUTO parameter (Bertin \& Arnouts 1996). We adopted for each object the 
same aperture among all passbands to ensure that the colors were measured correctly (Miyazaki et al. 2003; Furusawa et al. 2008).

The next step was to calculate the photometric redshifts (zphot) using the usual codes from the literature. In our case, we 
have made trials with several ones, and in the end we have selected the LePHARE program (Ilbert et al. 2006) based on 
a SEDs fitting method, because it gave the best results in comparison with available XMM-LSS objects spectroscopic 
redshifts. These photometric redshifts were computed with a large set of templates (Polletta et al. 2007), covering a 
broad domain in parameter space with a Calzetti \& Heckman (1999) extinction law. 

We have also compared our catalog with other zphot studies in the literature in the UDS field. Williams et al. (2009) have a large sample of objects with zphot calculated with the EAZY code (Brammer et al. 2008) and Rowan-Robinson et al. 2009 had used its own code, IMPZ code (Rowan-Robinson et al. 2002). All these codes have reasonable sources of error and this behavior is more evident at higher z. Therefore, we have estimated a standard deviation ($\sigma _z$) of four catalogs (in some sources only three redshifts are available): 
the Williams et al. catalog, the Rowan-Robinson et al. catalog and our results using apertures of 2" and 3" 
(our standard value in this paper). 

Among these sources, we consider only those for which $z>2.5$ that satisfy the selection criteria of Z-EROs/OGs
given in section \ref{.criteria} by the Eq. (\ref{criteria_eq}). 
There are not many available sources that comply with all of our constraints at $z>2.5$. We find only 45 galaxies, which decreases 
to 20 objects (see Fig. \ref{Fig:covered_area} and Table \ref{Tab:Candidates_zeros}) when we select only those for which 
$\sigma _z/(1+z)$ $\leqslant$ 0.1. That is, we reject the galaxies whose photometric redshift appears to depend on either 
the aperture size or the algorithm of photometric redshift determination. This is 
the sample that we use for our analysis to ensure that we are conservative in the redshift range 2.5 $< z <$ 3.8. 

As noted by L10, any compilation of EROs/OGs is not a homogeneous
sample of galaxies with similar intrinsic characteristics at all redshifts. The term 'Z-ERO' reflects the
observed characteristic colors of a galaxy, not its intrinsic properties, and the term 'OG' denotes that
the object is an old early-type galaxy rather than a starburst. In any case, the range of stellar
masses, M/L ratios, etc. may vary significantly among the 20 galaxies, and there may be selection effects. Nevertheless our 
analysis distinguishes the dependence on redshift from that on luminosity.
  
\begin{figure}
\begin{center}
\vspace{1cm}
\includegraphics[width=9.cm]{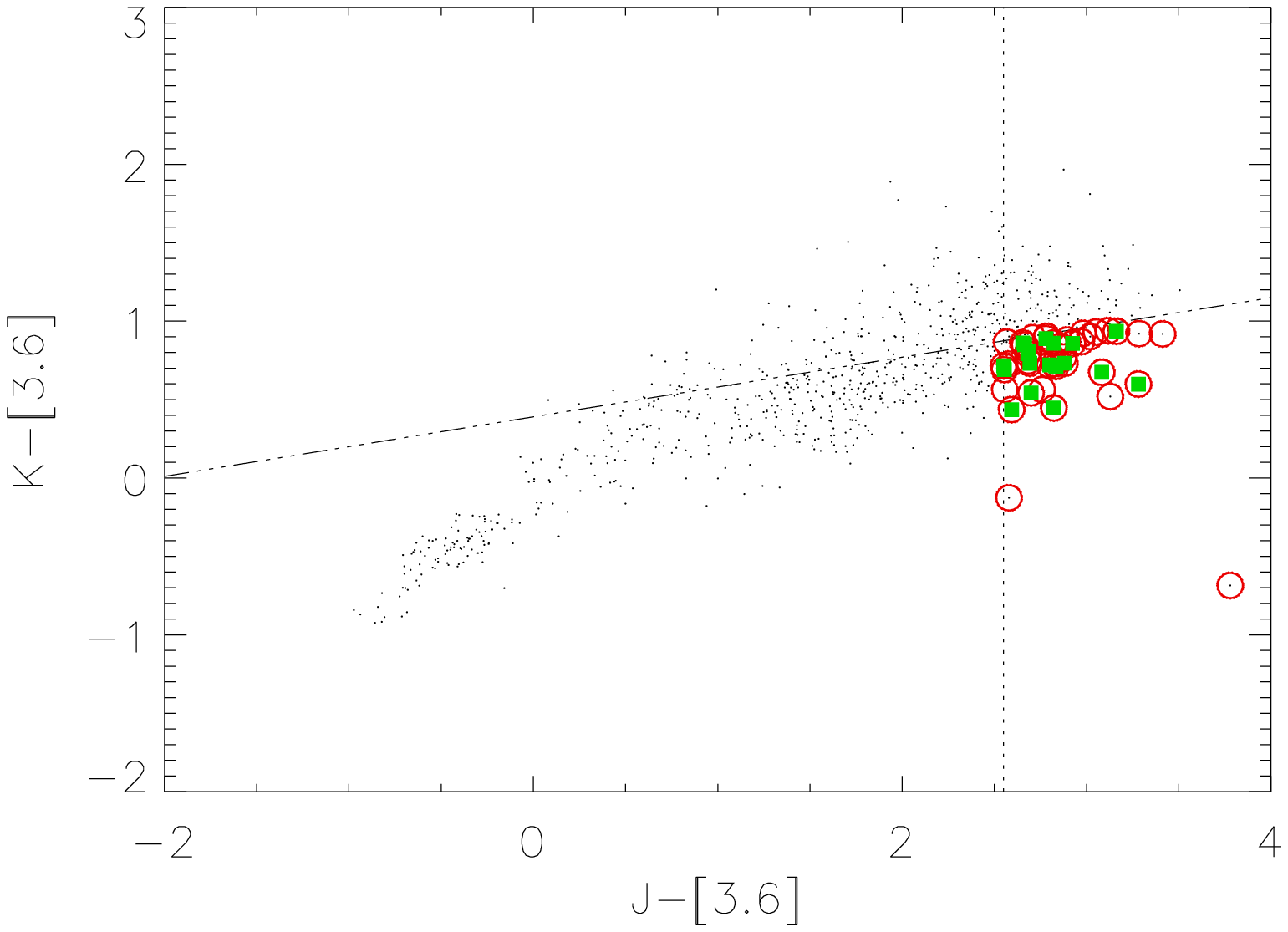}
\includegraphics[width=9.cm]{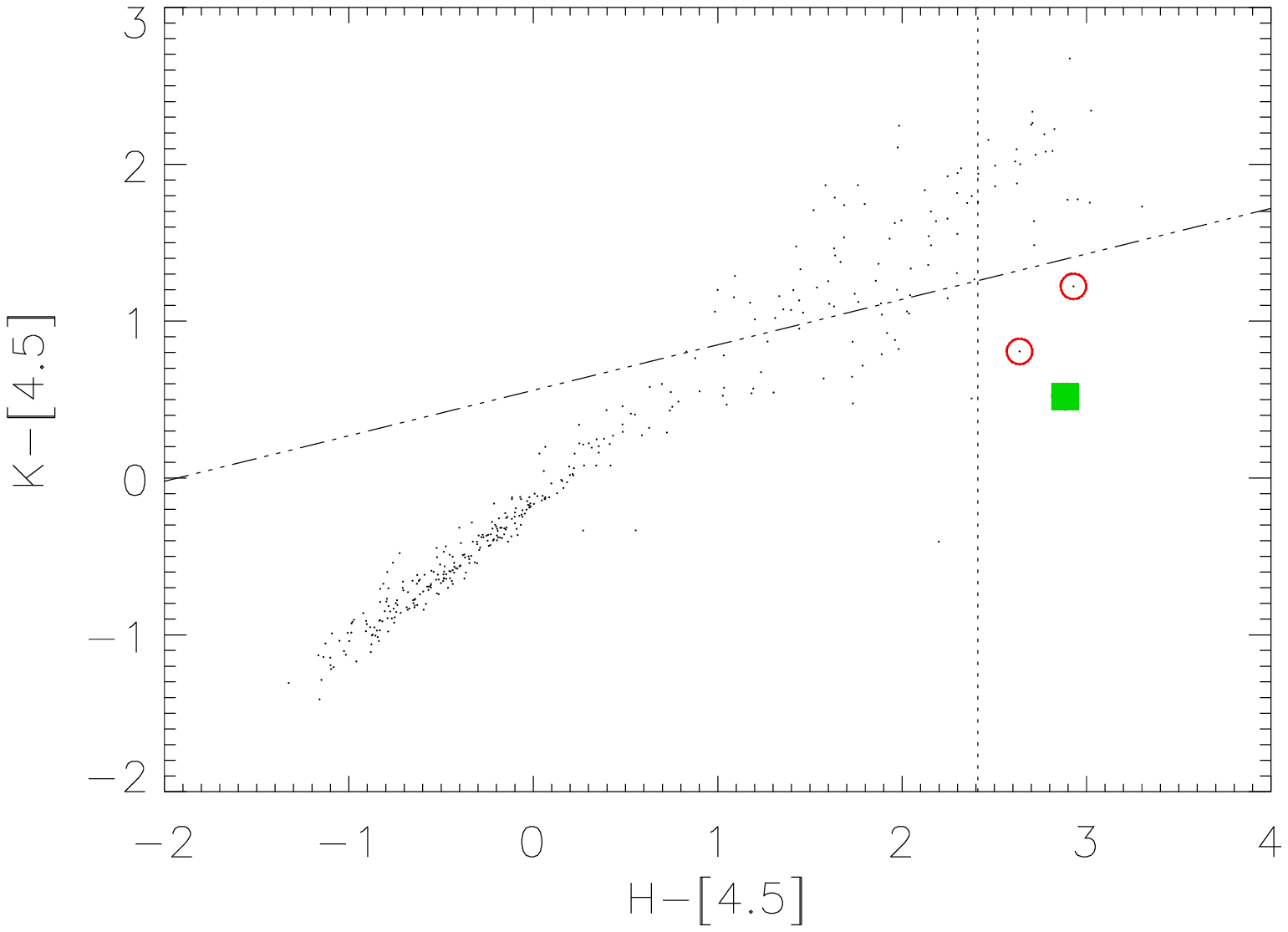}
\end{center}
\caption{Selection criteria used in this work to find the Z-EROs/OGs. a) Galaxies with
z between 2.5 and 3.5. b) Galaxies with z between 3.5 and 4.5. The red open circles represent the final 45 candidates to Z-EROs/OGs
following the Equation (\ref{criteria_eq}). The filled squares represent the final 20 selected galaxies with the zphot criterion.}
\label{Fig:equuatios_fig}
\end{figure}

\begin{table}
      \caption[]{General characteristics of the final 20 Z-EROs/OGs (objects with asterisks represent the galaxies with $\sigma _z/(1+z)$ $\leqslant$ 0.05.)}
         \label{Tab:Candidates_zeros}
   
         \begin{tabular}{c c c c c  c}
            \hline
            UDS Id & RA (deg.) & DEC (deg.) & z &(B-V)$_{rest}$ & t$_{obs}$ (Gyr) \\
            \hline
	    \hline
	    
06165  & 34.43191 &  -5.48688 &  2.511   &   0.786 & 2.553 \\
79102  & 34.14750 &  -4.77846 &  2.535   &   0.920 & 2.355 \\
25371  & 34.73086 &  -5.29274 &  2.536   &   0.834 & 2.238 \\
01559  & 34.40150 &  -5.53476 &  2.540   &   0.655 & 2.511 \\
40812* & 34.17666 &  -5.13515 &  2.565   &   0.771 & 2.716 \\
08303* & 34.29894 &  -5.46366 &  2.581   &   0.770 & 2.536 \\
24756  & 34.73077 &  -5.29849 &  2.668   &   0.797 & 2.665 \\
65143  & 34.40916 &  -4.91072 &  2.688   &   0.904 & 2.173 \\
78891  & 34.59163 &  -4.78038 &  2.705   &   0.583 & 2.360 \\
51547  & 34.90864 &  -5.03568 &  2.729   &   0.689 & 2.684 \\
39991* & 34.73174 &  -5.14295 &  2.754   &   0.888 & 2.574 \\
11488  & 34.64674 &  -5.43199 &  2.813   &   0.868 & 2.712 \\
85772  & 34.38530 &  -4.71782 &  2.852   &   0.583 & 2.487 \\
28153  & 34.28822 &  -5.26259 &  2.889   &   0.395 & 2.043 \\
28841* & 34.14779 &  -5.25500 &  2.896   &   0.798 & 2.716 \\
44569  & 34.42194 &  -5.10082 &  2.948   &   0.544 & 1.758 \\
10585  & 34.79973 &  -5.44024 &  3.033   &   1.060 & 2.394 \\
26888  & 34.24049 &  -5.27663 &  3.112   &   0.658 & 2.430 \\
14598* & 34.68966 &  -5.40086 &  3.287   &   0.680 & 2.744 \\
93574* & 34.28374 &  -4.63816 &  3.744   &   0.705 & 2.309 \\

            \hline
         \end{tabular}
    \end{table}

\begin{table*}
      \caption[]{The properties of the final 20 objects with $\sigma _z/(1+z)\leqslant 0.1$ subdivided into three bins of $N$ galaxies.}
         \label{Tab:Candidates_10bins}
         \begin{center}
         \begin{tabular}{c c c c c c c}
            \hline
 $\langle t_{\rm obs} \rangle $ (Gyr) & $N$ & $\langle (B-V)_{rest}\rangle $ & $\langle L_{V}\rangle $ ($10^{10}L\odot$) 
& $\langle M_*\rangle $ ($10^{10}M\odot$) & $\langle t_{\rm gal}\rangle $ (Gyr) & $\langle t_{\rm form}\rangle $ (Gyr) \\
            \hline
	    \hline
 2.18 & 7 &	$0.698 \pm 0.200$ & $14.0 \pm 4.9$ & 9.3	& 1.30 $^{+2.26}_{-0.52}$	& 0.88 $^{+0.52}_{-2.26}$		\\
 2.50 & 7 &	$0.772 \pm 0.163$ & $11.7 \pm 6.8$ & 11.3	& 1.82 $^{+3.67}_{-0.83}$	& 0.68 $^{+0.83}_{-3.67}$		\\
 2.70 & 6 &	$0.768 \pm 0.072$ & $8.0  \pm 6.5$ & 7.5	& 1.80 $^{+0.83}_{-0.62}$	& 0.90 $^{+0.62}_{-0.83}$		\\
            \hline
         \end{tabular}
       \end{center}   
    \end{table*}

\begin{table*}
      \caption[]{Properties of the six objects with $\sigma _z/(1+z)\leqslant 0.05$ (only one $t _{\rm obs}$ bin).}
         \label{Tab:Candidates_5bins}
   \begin{center}
         \begin{tabular}{c c c c c c c}
            \hline
         $\langle t_{\rm obs} \rangle $ (Gyr) & $N$ & $\langle (B-V)_{rest}\rangle $ & $\langle L_{V}\rangle $ ($10^{10}L\odot$) 
& $\langle M_*\rangle $ ($10^{10}M\odot$) & $\langle t_{\rm gal}\rangle $ (Gyr) & $\langle t_{\rm form}\rangle $ (Gyr) \\
            \hline
	    \hline
2.33 & 6 & $0.655 \pm 0.188$ & $16.5 \pm 5.5$ & 10.6 & 1.20 $^{+1.41}_{-0.48}$	& 1.13 $^{+0.48}_{-1.41}$		\\

            \hline
         \end{tabular}
        \end{center} 
    \end{table*}

\section{The rest-frame color $(B-V)$.}
\label{.colors}

We use AB apparent magnitudes that are corrected for Galactic extinction (assuming the average value for the field of A$_{V}$=0.07), for different wavelengths, $m_{AB}(\lambda _i)$, $(i=1,...,N_f)$,
for $5\le N_f\le 12$, with the corresponding error bars 
(in our case in the optical, near-infrared and mid-infrared, between B and [8.0]; we note that the [24.0] filter is not used here).
As we have already mentioned, we consider only 
data points with a flux S/N above 3. Applying the method of L10 (\S 3), we are able to determine the rest-frame color (B-V) of our selected galaxies.

\begin{figure}
\begin{center}
\vspace{1cm}
\includegraphics[width=9.cm]{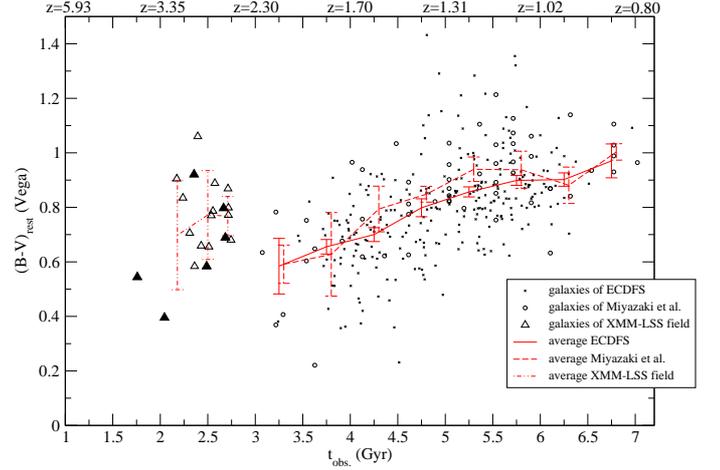}
\end{center}
\caption{$(B-V)_{\rm rest}$ (Vega calibration) colors observed
at different ages $t_{\rm obs.}$ (the age of the Universe corresponding to the given redshift). The XMM-LSS field are the data of this paper (the six filled triangles correspond to galaxies with $\frac{\Delta z}{1+z} \leqslant 0.05$), ECDFS, and ''Miyazaki et al.'' galaxies are taken from
L10.}
\label{Fig:colors}
\end{figure}

In Fig. \ref{Fig:colors}, we display the average colors
as a function of the age of the Universe when the galaxy is observed
\begin{equation}
t_{obs.}(z)=\frac{1}{H_0}\int _\infty ^z dx \frac{-1}{1+x}
\frac{1}{\sqrt{\Omega _m(1+x)^3+\Omega _\Lambda }}
\label{ageuniv}
.\end{equation}

In this figure, we have also added the galaxies at $z<2.5$ from L10.
The average gradient in color of the combined sample (including all galaxies in Fig. \ref{Fig:colors}, $0.8<z<3.8$) is $0.0877\pm
0.0093$ Gyr$^{-1}$.
For the galaxies at $z>2.5$ analyzed in this paper, the gradient is $0.20\pm
0.13$ Gyr$^{-1}$. As noted by L10, the gap in the galaxy data points in the lower right corner of Fig. \ref{Fig:colors} is at least partially an artifact of the sample
intrinsic color bias as a function of redshift. The step at $z\approx 2.5$ may be produced by the change in the color criterion to select EROs ($z\le 2.5$) or Z-EROS ($z>2.5$), bearing in mind
that our criteria for selecting Z-EROs emulates the colors of EROs at $z=1.5$ at higher redshifts (see \S \ref{.criteria}), hence we expect
to find that the average color of Z-EROs is similar to those of EROS at $z=1.5$.
Therefore, one must interpret Fig. \ref{Fig:colors} as a plot of the
intrinsic color versus (vs.) the redshift of the objects selected as (Z)-EROs/OGs, not as a general characterization 
of early-type galaxies.

If we perform a bi-linear fit of the colors as a function of two independent variables $t_{obs.}$ and $M_{V,rest}$,
we get 

\begin{equation}
(B-V)_{rest}=a_1+a_2[t_{obs.}({\rm Gyr})-2]
+a_3(M_{V,rest}+23)
\label{bmv2fit}
,\ \end{equation}
with $a_1=0.69\pm 0.06$, $a_2=-0.02\pm 0.14$, $a_3=0.14\pm 0.06$ for the 20 galaxies in this paper. Given the large error bars, we are unable to discern any gradient of colors with respect to age and metallicity.
L10 did measure significantly non-zero values of $a_2$ and $a_3$ at $z<2.5$.

\section{Age estimation for early-type galaxies}
\label{.age}

To estimate the average age and stellar mass of our galaxies, we 
apply the method described in L10 (\S 4). We assume a single stellar population for each galaxy; when there is instead a mixture of several populations, our ages would reflect an average of the age for the different populations. This method uses the synthesis model of Vazdekis et al. (1996), and breaks the age--metallicity degeneracy using the
mass--metallicity correlation.
We note that the role of thermally pulsing asymptotic giant branch (TP-AGB) stars is included in this synthesis model, and although some uncertainties are caused by 
this population (Maraston 2005), these are negligible for $\lambda < 6000 $ \AA \  (Bruzual 2007); these uncertainties are, however relevant for redder colors. Moreover, the effect of TP-AGB stars is important for ages younger than 1.5 Gyr and, as we later show, our galaxies are probably at this limit. For galaxies younger than 3 Gyr, the errors in metallicity are insignificant (L10), hence any evolution in the mass-metallicity relationship from high to low redshift would not significantly affect our results.

\begin{figure}
\begin{center}
\vspace{1cm}
\includegraphics[width=8.cm]{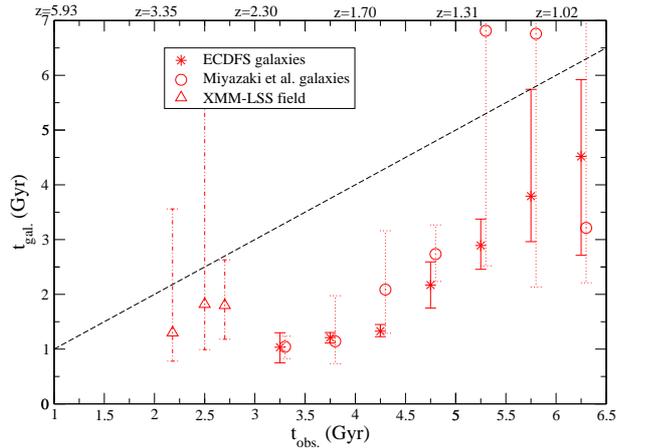}
\end{center}
\caption{Average age ($t_{gal}$) of the stellar populations of (Z-)EROs. The dashed line stands for the limiting maximum $t_{gal}=t_{\rm obs.}$ 
within the standard cosmology. The bin with $0.8<z<0.9$ is not plotted because it has a 
very large error bar (L10) and it does not provide any significant information.}
\label{Fig:ages}
\end{figure}

Although the method cannot be applied to each single galaxy separately, it can be used statistically,
by binning the galaxies within a small range of redshift and calculating the average age and its r.m.s. for the Z-EROs/OGs in each bin.
In our case, we divided our sample into three bins, each of which has an associated average intrinsic color (see Fig. 
\ref{Fig:colors} and Table \ref{Tab:Candidates_10bins}). The age of the early-type galaxies is plotted in Fig. \ref{Fig:ages}, together with the ages of the galaxies
with $z<2.5$ from L10. The bin with $0.8<z<0.9$ is not plotted in Fig. \ref{Fig:ages} because it has a very large error bar (L10) 
and it does not provide any significant information.
The vertical bars include the errors derived from the uncertainties in the intrinsic colors plus the uncertainties in the synthesis
model used in L10. We are neglect the systematic errors in the photometric redshifts; were they not negligible, we would have
additional systematic errors in the calculated ages. Figure \ref{Fig:ages} represents the average age
of the given sample among the (Z-)ERO/OGs, with all the selection effects associated with each redshift.

\begin{figure}
\begin{center}
\vspace{1cm}
\includegraphics[width=8.cm]{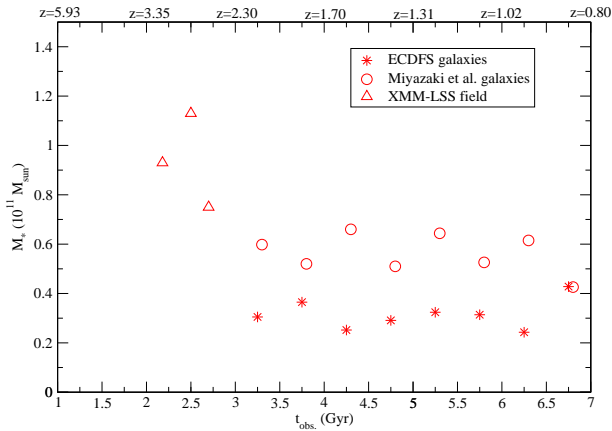}
\end{center}
\caption{Average stellar mass of (Z-)EROs which are passively evolving early-type galaxies.}
\label{Fig:masses}
\end{figure}

\begin{figure}
\begin{center}
\vspace{1cm}
\includegraphics[width=8.cm]{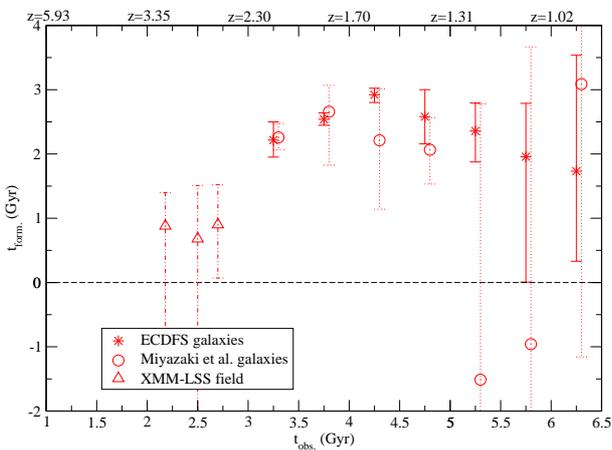}
\end{center}
\caption{Average age of the Universe at which the stellar populations of the
galaxies observed at age $t_{\rm obs.}$ have been formed.}
\label{Fig:ages_form}
\end{figure}

In Fig. \ref{Fig:masses}, we plot the stellar masses in each bin derived with the same code. The errors in the stellar mass  depend on both the data and variations in the initial mass function (IMF) in the Vazdekis et al. (1996) model, hence we can expect uncertainties of a factor two or even larger.  The stellar masses of the galaxies at $z>2.5$ are higher than 
$5\times 10^{10}$ $M_\odot $. We note that at very high redshift, the
stellar masses are much increased, as expected because of a selection effect that prevents us from detecting low
mass galaxies at those distances. To differentiate between the effects of the dependence on mass and evolution, we apply a 
bi--linear fit to the average color, weighted with the square inverse of the relative errors in the average age, including the $z<2.5$ bins of L10. We get

\begin{equation}
(B-V)_{rest}=b_1+b_2(t_{\rm obs.}-2)
+b_3\log_{10}(M_*)
\ ,\end{equation}
\begin{equation}
t_{\rm gal.}=c_1+c_2(t_{\rm obs.}-2)
+c_3\log_{10}(M_*)
\ ,\end{equation}
where both $t_{\rm gal.}$ and $t_{\rm obs.}$ are in units of Gyr, $M_*$ is in units
of $10^{11}\ {\rm M_\odot}$, and
$b_1=0.63\pm 0.04$, $b_2=0.096\pm 0.008$, $b_3=0.21\pm 0.06$,
$c_1=-0.18\pm 0.96$, $c_2=1.90\pm 0.29$, and  
$c_3=4.76\pm 1.77$.
The average epoch of star formation (the first 
stars might be form earlier) is 
$t_{\rm form.}=t_{\rm obs.}-t_{\rm gal.}$, 
as shown in Fig. \ref{Fig:ages_form}. Separating the evolution from the
mass dependence, 

\begin{equation}
t_{\rm form.}=d_1+d_2(t_{\rm obs.}-2)+
d_3\log_{10}(M_*)
,\end{equation}
where $d_1=2.18\pm 0.96$, $d_2=-0.91\pm 0.29$, and $d_3=-4.76\pm 1.77$.
For galaxies $0.8<z<2.5$ L10 had found that $d_1 = 1.94 \pm 0.51$, $d_2 = -0.46 \pm 0.32$, and $d_3 = -0.81 \pm 0.98$, 
hence with the extension of the redshift to 3.8 in this paper, we have got with respect to L10 measure values of $d_2$ and $d_3$ that differ significantly from zero. In L10, the stellar masses at $0.8<z<2.5$ were confined to a 
narrow range, although for the wider redshift range $0.8<z<3.8$ the range of masses is similarly much larger (see Fig. \ref{Fig:masses}), 
owing to Malmquist bias or other selection effects, permitting a more reliable determination of $d_3$.
The meaning of our numbers $d_2$ and $d_3$ is that, the lower the redshift (for a fixed mass), the lower the age of 
galaxy formation, and the higher the stellar mass (for a fixed redshift) the lower the age of galaxy formation. On 
average, for the whole range $0.8<z<3.8$, we find that 
$\left \langle \frac{\partial t_{\rm formation}}{\partial t_{\rm observed}}= -0.91\pm 0.29 \right \rangle$, 
$\left \langle \frac{\partial t_{\rm formation}}{\partial \log _{10}M_*}=-4.76\pm 1.77 \right \rangle $ Gyr,
which is significant at the 3.1$-\sigma $ and 2.7$-\sigma $ levels respectively.

\section{Discussion}
\label{.disc}

We have established a new color criteria for identifying objects of the same intrinsic colors (or even redder) 
than the EROs/OGs at redshift $0.8<z<2.5$ but at higher redshift (Z-EROs/OGs). This is useful for searching for the oldest galaxies at 
an epoch during which the Universe was younger than $\sim 3$ Gyr.

By adding 20 galaxies at redshifts $2.5<z<3.8$ to the catalog of L10, we have been able to see that we still derive formation ages that are within a narrow 
range (see Fig. \ref{Fig:ages_form}), of on average $2.0\pm 0.3$ Gyr ($z_{\rm form.}\approx 3-4$) for the galaxies 
observed at $0.8<z<2.5$ (L10), and
$0.9^{+0.4}_{-0.8}$ Gyr [simple weighted average of the three bins in Fig. \ref{Fig:ages_form}], and ($z_{\rm form.} \gtrsim 4.7$, 1$\sigma$) 
for the galaxies observed at $2.5<z<3.8$. Clearly, the new galaxies observed at higher
redshift were formed (around 1 Gyr) earlier, mainly because of the higher stellar mass.
This is the most astonishing result derived in this paper: that very massive galaxies were formed at redshifts 
$\gtrsim 4.7$.

If we wish to be conservative, we could analyze only the galaxies with $\sigma$z/(1+z) $\leqslant$ 0.05. In this case, we would have only six galaxies and only one bin in our calculations (see Table \ref{Tab:Candidates_5bins} and the black triangles in Fig. \ref{Fig:colors}). 
There is a slightly bluer average color and higher average luminosity, but still the galaxies are older than 1 Gyr on average,
and compatible with the results of the 20 galaxies within the error bars. Therefore, our results are quite insensitive to
the level of robustness of the redshift determination.

This agrees with the results mentioned in the introduction that very massive evolved galaxies 
detected at redshifts 1.5-6 were formed in the very early universe (Glazebrook et al. 2004, Daddi et al. 2005: Chen \& Marzke 2004; 
Rodighiero et al. 2007; Wiklind et al. 2008). Clearly, any hierarchical $\Lambda $-CDM model that concludes that very massive 
galaxies formed after the formation of low-mass galaxies similarly disagrees with these results.

\

{\bf Acknowledgments:}
We thank  A. Hempel (IAC) for helpful suggestions and comments on a draft of this paper. We thank to A. Vazdekis (IAC) for clarifications about his synthesis model (Vazdekis et al. 1996). Thanks are also given to the anonymous referee for helpful comments.
This work is based in part on data obtained as part of the UKIRT Infrared Deep Sky Survey; on observations 
with the Spitzer Space Telescope, which is operated by the Jet Propulsion Laboratory, California Institute of 
Technology under a contract with NASA; and on data collected at Subaru Telescope, which is operated by the National 
Astronomical Observatory of Japan.
NC is member of the Consolider-Ingenio 2010 Program grant CSD2006-00070: First Science with the GTC  
(http://www.iac.es/consolider-ingenio-gtc)" and the project EMIR AYA2009-06972 both of Spanish MICINN; and member of 
the SPIRE consortium led by Cardiff University.

\end{document}